# Geoantineutrino Spectrum and Slow Nuclear Burning on the Boundary of the Liquid and Solid Phases of the Earth's core

V.D. Rusov[1*], V.N. Pavlovich[2], V.N. Vaschenko[3-4], V.A. Tarasov[1], T.N. Zelentsova[1], D.A. Litvinov[1], S.I. Kosenko[1], V.N. Bolshakov[1], E.N. Khotyaintseva[2]

[1]*Odessa National Polytechnic University, 65044 Odessa, Ukraine*
[2] *Institute for Nuclear Researches of National Academy of Science, 01028 Kiev, Ukraine*
[3]*Ukrainian Antarctic Center, 01033 Kiev, Ukraine*
[4] *National Taras Shevchenko University, 01017 Kiev, Ukraine*

(February 2, 2004)

**Abstract**

The problem of the geoantineutrino deficit and the experimental results of the interaction of uranium dioxide and carbide with iron-nickel and silica-alumina melts at high pressure (5-10 GPa) and temperature (1600- $2200^0$ C) have induced us to consider the possible consequences of made by V. Anisichkin and A. Ershov supposition that there is an actinoid shell on boundary of liquid and solid phases of the Earth's core. We have shown that the activation of a natural nuclear reactor operating as the solitary waves of nuclear burning in $^{238}$U- and/or $^{232}$Th-medium (in particular, the neutron-fission progressive wave of Feoktistov and/or Teller-Ishikawa-Wood) such physical consequent can be.

The simplified model of the kinetics of accumulation and burnup in U-Pu fuel cycle of Feoktistov is developed. The results of the numerical simulation of neutron-fission wave in two-phase $UO_2$/Fe medium on a surface of the Earth's solid core are presented.

On the basis of O'Nions-Ivensen-Hamilton model of the geochemical evolution of mantle differentiation and the Earth's crust growth supplied by actinoid shell on the boundary of liquid and solid phases of the Earth's core as a nuclear energy source, the tentative estimation of intensity and geoantineutrino spectrum on the Earth surface was obtained.

---

[*] Corresponding author: Prof. Rusov V.D., E-mail: siiis@te.net.ua

# 1. Introduction

The problem of the geoantineutrino deficit [1] produces a need to consider the probability of the existence of additional energy sources in the interior of the Earth for renewal of geoantineutrino balance. Among a such sources there may be the actinoids, which are located lower than Gutenberg boundary or, in other words, lower than the mantle. We think that at present the experimental results of Anisichkin and Ershov [2] are the most developed mechanism of actinide shell formation lower than the mantle. According to this results the chemically stable and high-density compounds of actinoids (particularly carbides and uranium dioxides) almost completely lose their lithophile properties and could be lowered together with melted iron and concentrate in the Earth core during gravity differentiation of the planet substance. The concentration of actinides on the surface of the Earth's solid internal core could take place after gravity differentiation of substance, i.e. from 4 to $4.5 \cdot 10^9$ years ago. The hypothesis of actinoids concentration deep in the planet interior during gravity differentiation of substance were earlier expressed in Refs. [3-5]. The mechanism of actinides concentration in the Earth core is also of interest because it helps to solve the problem of an energy source for geodynamo operating, which creates the terrestrial magnetic field [4].

The self-propagating waves of nuclear burning in $^{238}$U- and/or $^{232}$Th-mediums must be the natural physical result of existing of such actinide shell in the Earth's core. In other words, in the thermal history of the Earth there must be some geophysical events, which will give a proof of the existence of the spontaneous reactor-like reactions of U-Pu and/or Th-U fuel cycles developed by Feoktistov [6] and Teller-Ishikava-Wood [7] respectively on the boundary of liquid and solid phases of the Earth's core. As it is shown below, a such geophysical events might be the isolated volcanoes, which are non-connected with the volcanism of plate boundaries, or so called hot spots, which formation mechanism meets Morgan's hypothesis of plumes, i.e. jets of hot substance escalating from the mantle depths (700-900 km) to lithosphere bottom [8].

The main purpose of the present paper is trial estimation of the intensity of the oscillating geoantineutrino flow on the Earth's surface from different radioactive sources ($^{238}$U, $^{232}$Th and $^{40}$K) by analysis of time evolution of radiogenic heat-evolution power of the Earth within the framework of the geochemical model of the mantle differentiation and the Earth's crust growth [8, 1], which is supplemented by the nuclear energy source located on the boundary of liquid and solid phases of the Earth's core.

## 2. The estimation of latent power of radiogenic heat-evolution of the Earth

Because of the Earth's high thermal inertia (low heat conduction) the heat, which is generated in the Earth's interior is transmitted to the surface not immediately, but is late for the thermal relaxation time of the Earth $\tau_E$, which is about $(1\div2)\cdot10^9$ years [8]. Consequently, the heat flow observed today (40 TW) carries the radiogenic heat, which was released a few billions years ago, when radioactive heat-evolution of the Earth was much higher than today. This result on quantitative level was obtained for the first time by O'Nionse, Ivensen and Hamilton in 1979 in their work devoted to the geochemical modeling of the mantle differentiation and the Earth's crust growth [8]. The calculated total radiogenic heat-evolution of the Earth in the context of the model of geochemical evolution of the Earth (from the initial composition to the modern composition of continental crust and external 50-km layer) with allowance for the content of dissipated elements in the Earth and based on available geo- and astro-chemistry data are given in Fig.1 [1]. Note that the selected value of uranium-thorium isotopic ratio, which took place 4.55 billions years ago (Fig.1), meets the data of $^{232}Th/^{238}U$ allowable values in *r*-process kinetic model [9], being the effective development of Fawler's theory of galactic nucleosynthesis [10].

Based on O'Nions-Ivensen-Hamilton geochemical model the geothermal flow power:

$$H(t=0) \cong 7.8(U) + 8.8(Th) + 2.9(K) = 19.5 \text{ TW}. \tag{1}$$

and integral antineutrino intensity on the Earth's surface, which was produced by natural uranium, thorium and potassium respectively:

$$\Phi_{\tilde{\nu}}(^{238}U) \approx 2.2\cdot10^6, \quad \Phi_{\tilde{\nu}}(^{232}Th) \approx 1.8\cdot10^6, \quad \Phi_{\tilde{\nu}}(^{40}K) \approx 9.12\cdot10^6 \quad \text{cm}^2\text{s}^{-1} \tag{2}$$

in Ref. [1] are obtained.

Let us try to estimate the value of additional geothermal flow due to actinoid shell, which was formed by chemical-stable and high-density compounds of actinoids. As is shown experimentally [2], these compounds could be lowered together with melted iron and concentrate on the surface of internal solid core ($r_N \approx 1200$ km) during gravity differentiation of substance.

Let us consider the structure and some behavior features of the power function of geothermal flow $H(t)$. It is known [8] that the heat flow from the core to the mantle (non-connected with radioactive decay) is approximately 1/7 of the total the Earth's heat flow $H$ (given as a dotted line in Fig.1). Following Ref. [8], we may suppose that the other component of the total Earth's heat flow $H$, which characterizes the initial heat originated during the formation of the planet and subsequent gravity differentiation on the core and the mantle, is also about 1/7 of $H$ (a dash-dotted line in Fig.1). Supposing that $\tau_E$ is about $\cdot10^9$ years, it is easy to show that the current, i.e. latent, power of radiogenic heat flow $H_L(t)$ is above the similar curve of $H(t)$ calculated by O'Nions-Ivensen-Hamilton [8] (see Fig.1), and now is

$$H_L(0) \cong 22.0 \quad TW. \tag{3}$$

Note that this value characterizes the new latent value of geoantineutrino flow intensity. Subtracting Eq. (1) from Eq. (3) we obtain the estimation of the additional power of the geothermal flow $H_f$, caused by the neutron- fission process of the actinoid shell of the Earth's solid core:

$$H_f = H_L(0) - H(0) = 2.5 \quad TW. \tag{4}$$

Obviously, if this power (4) is generated only owing to radiogenic heat, there will be no the contribution of actinide shell to the integral intensity of geoantineutrinos. In order to obtain the real contribution of actinide shell we suppose that the energy-release power $H_f$ of actiniod shell as a nuclear energy source is essentially higher than the partial power of radiogenic heat $H_\alpha$, produced by the radioactive chains of $^{238}$U and $^{232}$Th, i.e. $H_\alpha << H_f$.

Below, for simplicity sake, we consider the actinoid shell as the two-phase layer of $UO_2$/Fe on the surface of solid (iron) core of the Earth. Iron ($\rho \sim 12.0$ g/cm$^3$) in the pores of nuclear fuel ($\rho \sim 19.5$ g/cm$^3$), which typical "poured" concentration is $\sim 60\%$, decreases the density of the two-phase layer to $\sim 15$ g/cm$^3$. Let us assume $H_\alpha \sim 0.1$ TW. If the two-phase actinoid medium with the total mass of natural uranium

$$M(U) = H_\alpha / \varepsilon(U) \sim 10^{15} \quad kg, \quad where \quad \varepsilon(U) \cong 0.95 \cdot 10^{-4} \quad W/kg, \tag{5}$$

represents a continuous homogeneous shell on the surface of the Earth's solid core, its thickness will be less than 1 cm. Apparently, the image of a such two-phase actinoid medium as the inhomogeneous shell, which represents the stochastic web of "rivers and lakes" of actinoid medium, which is situated in the valleys of rough surface [11] of the Earth's solid core, is more correctly.

Below we show how the near-zero contribution of antineutrinos from the actinoid shell produced by $^{238}$U- and $^{232}$Th radioactive chains the total geoantineutrino flow on the Earth's surface can change qualitatively and quantitatively at the occurrence of self-propagating nuclear burning waves in $^{238}$U- and/or $^{232}$Th-medium. For that let us consider the simplified possible mechanisms of slow nuclear burning in the mediums, which emulate the operation of natural nuclear reactors.

### 3. The simulation of neutron-fission wave of Feoktisov

The mechanism of uranium concentration in the Earth core is in detail considered in Ref. [2]. The results of the experiments [2] on the interaction of uranium carbide and dioxide with nickel-iron and silica-alumina melts at high pressure (5÷10 GPa) and temperature (1600÷2200° C) give grounds to consider that on the early stages of the evolution of the Earth and other planets uranium and thorium oxides and carbides (as the most dense, refractory and marginally soluble at high pressures) could accumulate from the magma "ocean" on the solid internal core of the planet, thereby creating the possibility for the activation of chain nuclear reactions [2] and, particularly of Feoktisov [6] and/or Teller-Ishikawa-Wood [7] mechanism of progressing wave.

The geometric image of the natural "stationary" fast reactor, according to Ref. [6], could be pictured in the following way. Let there is an infinite cylinder of $^{238}$U about 1m in diameter. In some part of it there is a reaction focus formed forcedly, for example, due to enrichment by fissionable isotope. The next layers of uranium catch the neutrons escaping from reaction area and then $^{239}$Pu is efficiently produced in these layers. If the energy-release is sufficiently high, the concentration of $^{239}$Pu in adjoining areas become greater than the critical one and center of energy-release will shift. At the same time the accumulation of plutonium in next layers will begin. So, as result of a such fuel cycle (first proposed by Feoktistov in 1989 [6])

$$^{238}U(n,\gamma) \rightarrow {}^{239}U \xrightarrow{\beta} {}^{239}Np \xrightarrow{\beta} {}^{239}Pu(n, fission)... \qquad (6)$$

a progressing wave will arise, on front of which uranium is transformed to plutonium due to fission neutrons. In other words, neutron-fission wave transmission in $^{238}$U-medium is possible at a certain correlation between the equilibrium ($n_{Pu}$) and critical ($n_{crit}$) concentrations of plutonium, i.e. ($n_{crit} < n_{Pu}$). A wave velocity is about $L/\tau \sim 1.5$ cm/day (where $L \sim 5$ cm is diffusion distance of neutron in uranium and $\tau = 2.3/\ln2 = 3.3$ days is time of plutonium formation by β-decay of $^{239}$U). Note that besides delay time of neutrons one more time $\tau_{1/2} = 2.3$ days (which plays an important role in safety of Feoktistov natural reactor [6]) is appeared in scheme (6).

The similar idea is underlies of the mechanism of the formation of nuclear burning progressing wave in $^{232}$Th-medium corresponding to Teller-Ishikawa-Wood Th–U fuel cycle

$$^{232}Th(n,\gamma) \rightarrow {}^{233}Pa \rightarrow (\beta) \rightarrow {}^{233}U(n, fission)..., \qquad (7)$$

which was described in 1996 in Ref. [7].

In our paper the simplified model of Pu accumulation and U burnup kinetics is developed. In this model one-dimensional semi-infinite U-Pu medium irradiated from butt-end by external neutron source is considered in diffusion one-group approximation (neutron energy is ~ 1 MeV). The respective system of differential equations, which describes the kinetics of Feoktistov U-Pu fuel cycle taking into account delayed neutrons, i.e. the kinetics of initiation and transmission of neutron-fission wave $n(x, t)$, looks like:

$$\frac{\partial n(x,t)}{\partial t} = D \Delta n(x,t) + \left[ \frac{p\Phi_0 e^{-\chi(x,t)x}}{\chi(x,t)D \upsilon_n} + (1-p)n(x,t) + \sum_{i=1}^{6} \frac{\ln 2 \cdot \tilde{N}_i}{T^i_{1/2}} \right] F(x,t), \qquad (8)$$

where

$$F(x,t) = \upsilon_n \left[ (\nu - 1)\sigma_f^{Pu} N_{Pu}(x,t) - \sum_{i=8,\,9,\,Pu,\,f} \sigma_a^i N_i(x,t) \right],$$

$$\frac{\partial N_8(x,t)}{\partial t} = -\upsilon_n n(x,t) \sigma_a^8 N_8(x,t), \tag{9}$$

$$\frac{\partial N_9(x,t)}{\partial t} = \upsilon_n n(x,t) \sigma_a^8 N_8(x,t) - \frac{1}{\tau_\beta} N_9(x,t), \tag{10}$$

$$\frac{\partial N_{Pu}(x,t)}{\partial t} = \frac{1}{\tau_\beta} N_9(x,t) - \upsilon_n n(x,t)(\sigma_a^{Pu} + \sigma_f^{Pu}) N_{Pu}(x,t), \tag{11}$$

$$\frac{\partial \widetilde{N}_i}{\partial t} = p_i \cdot \upsilon_n \cdot n(x,t) \cdot \sigma_f^{Pu} \cdot N_{Pu}(x,t) - \frac{\ln 2 \cdot \widetilde{N}_i}{T_{1/2}^i}, \quad i = 1,6, \tag{12}$$

where $\widetilde{N}_i$ are the concentrations of neutron-rich fission fragments of $^{239}$Pu; $\Phi(x,t) = \frac{\Phi_0}{\chi(x,t)\widetilde{D}} e^{-\chi(x,t)x}$ is neutron-flux density from external source [12], $\Phi_0$ - is neutron-flux density from plane external source located on the boundary at $x=0$, $\chi(x,t) = \left[ \frac{\Sigma_a(x,t) + \Sigma_f(x,t)}{\widetilde{D}} \right]^{\frac{1}{2}}$; $\Sigma_a(x,t) = \sum_{i=8,\,9,\,Pu,\,f} \sigma_a^i N_i(x,t)$ is neutron-capture macro-cross-section; $\Sigma_f(x,t) = \sum_{i=8,\,9,\,Pu,\,f} \sigma_f^i N_i(x,t)$ is fission macro-cross-section; $\widetilde{D} = D/\upsilon_n$; $D$ is diffusion constant of a neutron; $\upsilon_n$ is neutron velocity ($E_n = 1$ MeV).

Deriving Eq. (8) we are supposed that $n(x,t) = q(x,t) + \Phi(x,t)\frac{1}{\upsilon_n}$, where $q(x,t)$ is fission neutron-flux density, which obeys Fermi expression [13]:

$$q(x,t) = n(x,t)(1-p) + \int_0^\infty n(x,t-t') \sum_{i=1}^6 \frac{\ln 2 \cdot p_i}{T_{1/2}^i} e^{-\ln 2 \cdot t'/T_{1/2}^i} dt', \tag{13}$$

where $p_i$ ($p = \sum_{i=1}^6 p_i$) are the parameters characterizing the groups of delayed neutrons for main fuel fissionable nuclides [12].

The boundary conditions for the system of differential equations (8)-(12) are

$$n(x,t)\big|_{x=0} = \frac{\Phi_0}{\chi(0,t)D}, \tag{14}$$

$$-D\frac{\partial n(x,t)}{\partial x}\bigg|_{x=0} = -D\frac{\partial}{\partial x}\left\{\frac{1}{D\chi(x,t)}\Phi_0 e^{-\chi(x,t)x} + q(x,t)\right\}\bigg|_{x=0}. \quad (15)$$

The initial conditions for the system of differential equations (8)-(12) are

$$n(x,t)\big|_{t=0} = \frac{\Phi_0 e^{-\chi(x,0)x}}{D\chi(x,0)} + q(x,t)\big|_{t=0}, \quad (16)$$

$$N_8(x,t)\big|_{t=0} = \frac{\rho_8}{\mu_8}N_A \approx \frac{19}{238}N_A,$$

$$N_9(x,t)\big|_{t=0} = 0, \quad N_{Pu}(x,t)\big|_{t=0} = 0, \quad \Sigma_a(x,t)\big|_{t=0} \approx \sigma_a^8 N_8(x,t)\big|_{t=0}, \quad \widetilde{N}_i^{Pu}(x,t) = 0, i=1,...,6;$$

$$D = \frac{\upsilon_n}{3\cdot\sigma_s^8 N_8(x,0)}; \quad \chi(x,0) = \left[\frac{\Sigma_a(x,t)\big|_{t=0} + \Sigma_f(x,t)\big|_{t=0}}{\widetilde{D}}\right]^{1/2}, \quad (17)$$

where $\rho_8$ - is the density, which is expressed in units of g·cm$^{-3}$; $N_A$- Avogadro constant.

The estimation of neutron flux density of an internal source on the boundary $\Phi_0$ can be obtained reasoning from the estimation of plutonium critical concentration ~ 10%:

$$4\tau_\beta \Phi_0 \sigma_a^8 N_8(x,t)\big|_{t=0} = 0.1 N_8(x,t)\big|_{t=0} \quad (18)$$

and so

$$\Phi_0 \approx 0.1/4\tau_\beta \sigma_a^8. \quad (19)$$

The following values of constants for simulation was used:

$\sigma_f^{Pu} = 2.0\cdot 10^{-24}$ cm$^2$; $\sigma_f^8 = 0.55\cdot 10^{-24}$ cm$^2$; $\sigma_s^8 = 2.3\cdot 10^{-24}$ cm$^2$; $\sigma_a^8 = 5.38\cdot 10^{-26}$ cm$^2$;

$\sigma_a^9 = 2.12\cdot 10^{-26}$ cm$^2$; $\sigma_a^{Pu} = 2.12\cdot 10^{-26}$ cm$^2$; $\nu = 2.9$; $\tau_\beta \approx 3.3$ days; $\upsilon_n \approx 10^9$ cm/s; D $\approx 10^9$ cm$^2$/s.

The program for the solution of the system of equations (8)-(12) taking into account boundary (14)-(15) and initial (16)-(17) conditions was made for Microsoft Power Station 4.0. The obtained results are presented in Figs. 2-3. The value $\Phi_0$=1.9·10$^{24}$ cm$^{-2}$s$^{-1}$ (Fig. 2) is obtained from expression (19) at $\tau_\beta \sim 3,3$ days. At the same time to reduce the calculating time for the solving of the system of equations (8)-(12) we used $\tau_\beta$=1 s.

Obviously, the numerical solution of the system of equations (8)-(12) with different parameters confirms the fact of originating self-regulating neutron-fission wave (the calculations for real geometry, multigroup approximation for a neutron spectrum and heat transmission equations will be considered in next paper).

As nuclear energy-release is high, a considerable warming-up takes place at quite small depth of reaction. In this case the heat sink is lightened by the low velocity of neutron-fission wave and is realized by the liquid-metallic coolant (iron), which is present in the area of actinoid shell on the border of the solid and liquid phases of Earth's core. Let us consider the nuclear-geophysical aspects of the initiation of the progressing wave of nuclear burning in real $^{238}$U-medium.

Two-phase layer $UO_2$/Fe on the surface of Earth's solid core is a natural medium for neutron-fission wave development. Since in a such wave contemporary and even depleted uranium can react, let us estimate the real possibility of wave process. The critical concentration of pure $^{239}$Pu in $^{238}$U in infinite medium, which was calculated by octa-group constants, is about 3.7 % [14]. Dilution by oxygen ($UO_2$/$PuO_2$) leads to the increase of critical concentration to $n_{crit}$ ~ 6.4 %. The presence of iron in nuclear fuel pores (with typical "poured" concentration about 60 %) will increase the critical concentration of $^{239}$Pu up to $n_{crit}$ ~ 8.2 % (for calculation $\rho$ ~ 19.5 g/cm$^3$ for $UO_2$/$PuO_2$ and $\rho$ ~ 12 g/cm$^3$ for Fe were used) [15]. Non-trivial thermodynamics conditions, i.e. high temperature and pressure, might rise the critical concentration of Pu up to $n_{crit}$~10 %. This means that the model system of equations (8)-(12) qualitatively close reflects the main properties of real breeding medium, taking into account that the addition of oxygen and Fe practically does not change the solutions because their neutron-absorption cross-sections are less, at least, by the order than the similar values for actinides.

Let us crudely estimate the diameter $D_0$ of Feoktistov's cylindrical uranium-plutonium reactor with power $P$ = 2.5 TW. Let axis of symmetry $Y$ of such a reactor is wave line of nuclear burning. From the kinetic dependencies for $^{239}$Pu given in Fig. 2 and Fig. 3 it is possible to estimate the width of burning wave $\Delta Y$ at steady-state regime. Let us accept it equal to 2 cm ( Fig. 2) and 10 cm (Fig. 3). As the volume $V$ of burning area at steady-state regime practically does not change, the fission rate of $^{239}$Pu $\eta_f$ in this area is:

$$\eta_f(^{239}Pu) \approx \Phi \cdot \sigma_f \cdot C_0 \cdot V = \Phi \cdot \sigma_f \cdot C_0 \cdot \pi R^2 \cdot \Delta Y, \quad (20)$$

where $\Phi = v_n \cdot C_n$ is neutron-flux density; $v$ is the neutron velocity; $C_n$ is the neutron concentration, $\sigma_f$ is fission cross-section for $^{239}$Pu; $C_0$ is the concentration of $^{239}$Pu.

Then we obtain the estimation of the cylindrical reactor diameter, which is characterized by U-Pu kinetics (6):

$$D_0 \approx 2\left(P/E_f \cdot v \cdot C_n \cdot \sigma_f \cdot C_0 \cdot \pi \Delta Y\right)^{1/2} \approx \begin{cases} 1.0 \ cm & for \ fig.2 \\ 200 \ cm & for \ fig.3 \end{cases}, \quad (21)$$

where $\eta_f = P/E_f$; $P$ = 2.5 TW; $E_f$ = 210.3 MeV is the fission energy of $^{239}$Pu; $v \approx 10^9$ cm·s$^{-1}$ is neutron velocity for energy 1 MeV; $\sigma_f \approx 2.0 \cdot 10^{-24}$ cm$^2$; the values of $\Delta Y$ = 2 cm, $C_n \approx 4 \cdot 10^{16}$ neutron·cm$^{-3}$, $C_0$ $\approx 10^{21}$ nucleus·cm$^{-3}$ and $\Delta Y = 10$ cm, $C_n \approx 2 \cdot 10^{11}$ neutron·cm$^{-3}$, $C_0 \approx 5 \cdot 10^{20}$ nucleus·cm$^{-3}$ are obtained by analyzing of model reactor kinetics, which is presented in Fig. 2 and Fig. 3 respectively. Note that the value $D_0 \approx 1.0 \div 200$ cm of Feoktistov's cylindrical uranium-plutonium reactor with power rating $P = 2.5$ TW well agrees with the probable thickness of $UO_2$/Fe two-phase medium, which, as shown above, represents a stochastic web of "actinoid rivers and lakes" situated in the valleys of rough (by an altitude up to 10-20 km) surface [18] of the Earth's solid core.

In spite of the active discussions of the possibility of chain nuclear reaction existence in interior of the Earth and other planets in numerous papers (starting with Kuroda [16] and ending with Driscoll [3], Herndon [4], Ershov and Anisichkin [2, 5, 15]), the question of the natural external neutron sources, which locally start the mechanism of nuclear burning, remains open and requires serious joint efforts of the theorists.

However, taking into account all difficulties concerning the explanation of the mechanism of neutron-fission wave starting, it is possible to go by alternative route and to try to find in the thermal history of the Earth geophysical events, which directly or indirectly denote the existence of slow nuclear burning. Note that these events should be in recent times, which as the present, characterized by lowered, i.e. subcritical concentration of odd isotopes of uranium and plutonium. Let us consider below the example of such geophysical paleo-events.

### 4. The hot spots statistics and the processes of slow nuclear burning

Along with the tectonics of plates that is the surface manifestation of convective movements in mantle interior, not less bright "traces" of convection in the mantle are detected on the surface of the Earth. These are isolated volcanoes or, according to Morgan [17], hot spots, which play a significant role in the understanding of the connection between tectonics of plates and the hydrodynamics of the mantle. The interest towards the hot spots arose in 1972 after Morgan proposed hypothesis of plumes, i.e. the hot jets in the mantle, which reach the basis of the lithosphere. In such points the lithosphere locally rises a little and melts. On the surface of the Earth the diameter of such hot spots (the cupola-shaped structures formed by outflow of basalt magma) can rich 200 km. According to Burke and Wilson [8], who first determined spatial hot spots distribution on the lithosphere plates, on the Earth's surface it is possible to detect at least 122 hot spots (Fig. 4), which were magma-active for last 10 millions years.

From the position of new global tectonics the scenario of hot spot formation is quiet simple and natural. It is supposed that in the Earth's mantle there are three boundary layers, i.e. three zones of decreased viscosity. The first zone is the classical asthenosphere of the Earth. The second zone of decreased viscosity is situated at the depth about 700-900 km. The third asthenosphere zone of the mantle is situated at its bottom (~ 2600−2885 km) on the boundary with the Earth core. Due to gravity instability a "piece" of overheated substance from the third boundary layer with the horizontal scale equal to its thickness (~200-300 km) slowly goes up and leaves a layer, and "cold" material of the overlying mantle takes it's place[*]. The situation may be repeated once again in the second boundary layer. If the second boundary layer is strongly overheated (for example, due to the "hot piece" of overheated substance from the third layer) it can become gravity unstable. Then the "pieces" of this layer with the horizontal scale equal to its thickness ~ 150 km will begin "to break away" from it and to float up to the bottom of the lithosphere, initializing hot spots formation on the surface of the Earth. The overheat mechanism of boundary layers is still unknown.

We assume, that Morgan plumes (which have energy threshold) are the visual geophysical "traces" of natural reactor operation in actinoid shell on the boundary of the liquid and solid phases of the Earth's core. In other words, the third boundary layer overheating is caused by energy-release of Feoktistov neutron-fission progressing wave. Below we shall try to present the arguments in favour of such a supposition basing on statistics of hot spots on the surface of the Earth.

Let the statistics of hot spots is formed by two-cascade stochastic process, where the primary random process (the number of progressing waves $v$) generates the secondary random process (the number of plumes per one progressing wave). In this case the secondary random process happens on the characteristic interval of length $r$ like shot noise [18]. Let us assume that both processes are Poisson. At the same time we consider the secondary process as inhomogeneous Poisson process characterized by velocity or so-called response function $h(x)$ [19]. Well-known spatial distribution of hot spots on the Earth's surface (Fig. 3) [8] is broken into a grid of squares with $d$ on side (where $d$ is size of a square detector's window [20]). Then the number of hot spots (or plumes), which fall into detection interval $[0, d]$, is counted. In this case the statistics of hot spots is described by so-called double-stochastic Poisson point process controlled by response function. At such formulation the determination of the type of hot spots statistics is reduced to well-known Neyman-Scott model [21], which in case of rectangular and exponential response function $h(x)$ was investigated by Saleh and Teich [22], and, when the response function has power kind, it was investigated by Lowen and Teich [23].

The generating function of double-stochastic Poisson point process controlled by response function $h(t)$ looks like

$$\Phi(z) = \exp\left[\mu \int_{-\infty}^{\infty} \{\exp(-(z-1)h_d(x)) - 1\} dx\right], \qquad (22)$$

---

[*] According to Ref. [8], the temperature of cold material from second boundary layer is 300-500° K lower than temperature of the third boundary layer. Such thermal perturbation in the third boundary layer harshly changes the hydrodynamic currents in the core and, perhaps, can change polarity of geomagnetic dipole of the Earth.

where $\mu$ is the intensity of first cascade, $h_T(x)$ is the convolution of response function $h(x)$ on interval $[0, d]$:

$$h_d(x) = \int_0^d h(x + x')dx'. \qquad (23)$$

The probabilities $p(n)$, which are the expansion coefficients $\Phi(z)$ in a Maclaurin's series, can be determined by a generating function:

$$p(n) = \frac{1}{n!} \cdot \frac{d^n \Phi(z)}{dz^n}\bigg|_{z=0}, \quad n = 0,1,2,... \qquad (24)$$

Using (22) and (24) it is possible to show that the recurrence relations for the distribution $p(n)$ of event number (hot spots) have a form:

$$(n+1)p(n+1) = \sum_{k=o}^{n} C_k p(n-k), \qquad (25)$$

$$p(0) = \exp\left\{\mu \int_{-\infty}^{\infty} \{\exp[-h_d(x)] - 1\}dx\right\} \qquad (26)$$

where

$$C_k = \frac{\mu}{k!} \int_{-\infty}^{\infty} [h_d(x)]^{k+1} \exp\{-h_d(x)\}dx. \qquad (27)$$

The response function normalized to the average number of hot spots per one Feoktistov neutron-fission progressing wave has a form:

$$\langle \varepsilon \rangle = \int_0^{\infty} h(K,x)dx, \qquad (28)$$

Then for exponential response function describing the decrease rate of energy release power in front of progressing nuclear burning wave

$$h(x) = \frac{\langle \varepsilon \rangle}{x_0} \exp\left(-\frac{x}{x_0}\right) = \frac{2\langle \varepsilon \rangle}{r} \exp\left(-\frac{2x}{r}\right), \quad r = 2x_0 \qquad (29)$$

expressions for average $\langle n \rangle$, dispersion $var(n)$ and third central momentum $\langle \Delta n^3 \rangle$ look like:

$$\langle n \rangle = \langle v \rangle \langle \varepsilon \rangle, \qquad (30)$$

$$var(n) = \langle n \rangle \left[1 + \frac{\langle \varepsilon \rangle r}{2d}\left\{\exp\left(-2\frac{d}{r}\right) + 2\frac{d}{r} - 1\right\}\right], \qquad (31)$$

$$\langle \Delta n^3 \rangle = \langle n \rangle \left[ 3 \frac{\text{var}(n)}{\langle n \rangle} - 2 + \frac{\langle \varepsilon \rangle r}{4d} \left\{ 4 \exp\left(-2\frac{d}{r}\right) - \exp\left(-4\frac{d}{r}\right) + 4\frac{d}{r} - 3 \right\} \right]. \quad (32)$$

where $<\nu> = \mu d$ is the average number of random events of the first cascade, or, in our case, the number of hot spots got to the detector with square window $d$ on side.

One of the ways of the identification and quantitative analysis of experimental distribution parameters lies in $\chi 2$–comparison with the theoretical distribution having exponential response function (see Eqs. (25)-(26)) and adds up to the determination of the parametric triplet $\{<\nu>, <\varepsilon>, d/r\}$, which is the solution of the system of nonlinear equations (30)-(32).

The procedure of the determination of distribution (25)-(26) parameters ($<\nu>$=0.64, $<\varepsilon>$=0.75, $d/r$=7.0) allowed to show the fitting high quality ($\chi^2$/NDF~2.7/2) of the experimental distribution (with averaged parameter $d \approx 2011$ km) of hot spots formed on the Earth's surface during last 10 millions years (Fig. 6).

The relevancy of the interpretation of the first cascade as random process of progressing waves formation rather than, for example, the large fluctuations of temperature is a natural question. Thereupon we give the estimations, which indirectly confirm the capability of low nuclear burning.

As stated above, the speed of Feoktistov progressing wave is about $L/\tau \sim 1.5$ cm/day (where $L$~5.0 cm is diffusion distance of neutron in $UO_2$/Fe). The analysis of the two-cascade statistics of hot spots has shown that $d/r$ =7.0. Hence, the length of characteristic interval of progressing wave, during which secondary events (Morgan plumes) are originated, is equal approximately $r = d/7.0$ ≈300 km (taking into account the projection averaging $d \approx$2011 km (Fig. 4)). Then rough but independent approximated estimation of Morgan plume number produced during last 10 millions years (on condition that one progressing neutron-fission wave always exists) is equal

$$N = \frac{t_0}{r/(L/\tau)} \cdot \langle \varepsilon \rangle = \frac{10^7}{3 \cdot 10^7 / 1.5 \cdot 365.25} \cdot 0.75 \approx 137. \quad (33)$$

This estimation satisfactorily agrees with the calculations of Burke and Wilson [8], who have shown that there are at least 122 hot spots on the Earth's surface, which were magma-active during last 10 millions years.

Note that hot spots observed last 10 millions years are produced by Morgan plumes, which, in its turn, were formed in the bottom boundary layers of the mantle approximately $10^9$ years ago. This time is in proportional to the thermal relaxation time of the Earth. Thus, if the plumes are result of self-propagating Feoktistov nuclear burning waves, they, at best, are only the modern witnesses of bright, but past deep-focus nuclear "catastrophes".

## 5. The integral intensity and geoantineutrino spectrum

The fission rate of $^{239}$Pu in neutron-fission wave front is

$$\eta_f = P/E_f \approx 7.4 \cdot 10^{-22} \text{ fission/s}, \qquad (34)$$

where $E_f$ is the average energy per fission of $^{239}$Pu.

Hence, the crude estimation of integral antineutrino intensity in two diametrically opposite points on the surface of the Earth from the front of burning wave in actinides web formed of $UO_2$/Fe has a form:

$$\Phi_{\tilde{\nu}} = \frac{1}{4\pi(R_\oplus \pm r_N)^2} \cdot \eta_f \cdot \mu_{\tilde{\nu}} = \begin{cases} 0.06 \cdot 10^6 \, cm^{-2}c^{-1}, & \text{if } "+", \\ 0.13 \cdot 10^6 \, cm^{-2}c^{-1}, & \text{if } "-". \end{cases} \qquad (35)$$

where $\mu_{\tilde{\nu}} \approx 5.7$ is the number of antineutrinos per fission of $^{239}$Pu; $R_\oplus \approx 6400$ km; $r_N \approx 1200$ km.

Using the design procedure of partial and total energy $\beta$-, $\tilde{\nu}$-spectra of radioactive nuclides [1] we have constructed the partial $d\Phi_{\tilde{\nu}}/dE$ ($^{238}$U), $d\Phi_{\tilde{\nu}}/dE$ ($^{232}$Th), $d\Phi_{\tilde{\nu}}/dE$ ($^{40}$K) [1], $d\Phi_{\tilde{\nu}}/dE$ ($^{239}$Pu) (Fig. 6) and the total energy antineutrino spectra of the Earth $d\Phi_{\tilde{\nu}}/dE$ ($^{238}$U+$^{232}$Th+$^{40}$K+$^{239}$Pu). The partial contributions were previously normalized to corresponding geoantineutrino integral intensity on the Earth's surface [1]).

The theoretical form of measured total energy spectrum $dn_{\tilde{\nu}}/dE$ without taking oscillations (Fig. 7) looks like

$$\frac{dn_{\tilde{\nu}}}{dE} = \varepsilon N_p \sum_i \frac{d\lambda_{\tilde{\nu}}^i}{dE} \sigma_{vp}(E_{\tilde{\nu}})\Delta t, \quad \text{MeV}^{-1} \qquad (36)$$

where $d\lambda_{\tilde{\nu}}/dE \equiv d\Phi_{\tilde{\nu}}/dE$ at $E_{\tilde{\nu}} \geq 1.804$ MeV; according to Ref. [11] $\varepsilon \approx 0.783$ is detector efficiency; $N_P = 3.46 \cdot 10^{31}$ is proton number in the detector sensitive volume; $\Delta t = 1.25 \cdot 10^7$ s is exposure time; $\sigma_{vp}$ is antineutrino-proton interaction cross-section for the inverse $\beta$-decay reaction with the allowance for corresponding radiation corrections [29-31].

The quantitative estimation of KamLAND-experiment data obtained by the integral of total spectrum (29):

$$n_{\tilde{\nu}} = \varepsilon \cdot N_p \cdot \Delta t \cdot \int_{E=1.804}^{\infty} \frac{d\lambda_{\tilde{\nu}}(U+Th+Pu)}{dE} \cdot \sigma_{vp}(E)dE =$$

$$= 3.48(U+Th) + 1.13(Pu)|_{E \leq 3.272} + 2.67(Pu)|_{E > 3.272} = 7.28 \qquad (37)$$

shows that in energy area $E \in 1.804 \div 3.272$ MeV without taking oscillations it is 2 times less and taking into account neutrino oscillations (under the condition of transition probability $P_{\tilde{\nu} \to \tilde{\nu}} \cong 0.55$ [32]) it is approximately 3.5 times less than the similar integral of best fit of reactor antineutrino experimental KamLAND spectrum obtained on the assumption of 9 geoantineutrinos detection [32].

In the future, when the KamLAND-experiment will have a sufficient statistics, the behavior of antineutrino spectrum in energy area $E \in 1.804 \div 3.272$ MeV can indirectly denote the existence of fission geoantineutrinos. This will require $\chi^2$-analysis of reactor antineutrino spectrum in the similar way as it was made in Ref. [33], but taking into account the fission geoneutrino background.

In any case the contribution of geoantineutrinos from a hypothetical nuclear reactor with 2.5 TW capability to total antineutrino spectrum of the Earth is significant and exerts influence on the energy structure of total antineutrino spectrum (Fig. 6). The future experiments should reveal the relevant singularities of the statistics, which will uniquely specify if fission geoneutrinos are necessary for the nature. May be the problem of oscillating geoneutrino deficit does not exist at all, and there is the problem of the statistics of observation data and the reliability of error of control energy antineutrino spectrum in KamLAND-like experiments.

## 6. Discussion and conclusions

On the basis of the temporal evolution analysis of radiogenic heat-evolution power of the Earth within the framework of the model of geochemical processes of the mantle differentiation and the Earth's crust growth [1, 8] supplied by a nuclear energy source on the boundary of the solid and liquid phases of the Earth's core we have obtained the tentative estimation of intensity and geoantineutrino spectrum on the Earth surface from different radioactive sources ($^{238}$U, $^{232}$Th, $^{40}$K and $^{239}$Pu).

We have also showed that natural nuclear reactors may exist on the boundary of the solid and liquid phases of the Earth's core as spontaneous reactor-like processes with U−Pu and/or Th−U Feoktistov [6] and Teller-Ishikawa-Wood [7] fuel cycle respectively.

The solution of the main problem connected with the search of natural neutron sources, which locally start the mechanism of nuclear burning, is unclear and (in spite of the active discussions of the possibility of the existence of chain nuclear reaction in interior of the Earth and other planets in the numerous papers [2-5, 15,16]) requires the serious joint efforts of the theorists.

However, taking into account all difficulties concerning the explanation of the mechanism of neutron-fission wave starting, it is possible to go by an alternative route and to try to find in the thermal, seismic or magnetic history of the Earth such geophysical events, which directly or indirectly denote the existence of slow nuclear burning. These events took place in recent times, which as nowdays are characterized by the lowered, i.e. subcritical concentration of odd isotopes of uranium and plutonium. First of all it concerns Morgan plumes, energy sources and geomagnetic field inversion, anomalous high (almost "solar") isotope ratio $^3$He/$^4$He=3·10$^{-5}$ in the mantle [34] etc. The spatial-temporal evolution of these events evidently reflects the changes of the mode and/or intensity of the natural nuclear reactors operating on the surface of the Earth's solid core.

As we indicated above, the statistics of hot spots on the Earth's surface is satisfactory explained by the natural nuclear reactor existence in the form of Feoktistov's nuclear burning progressing waves.

Now let us discuss the problem of energy sources and the Earth's geomagnetic field inversions. Under the theory of the Earth's hydromagnetic dynamo [8], the magnetic field is generated by electric currents in conductive the Earth's core. So, the available capacity of the external energy source for maintenance of hydromagnetic dynamo operating must at least compensate Joule energy loss (about 0.5 TW) caused by currents in the liquid core. It is obvious that the heat-evolution power, which is about 2.5 TW, in Feoktistov neutron-fission wave front exceeds with interest the energy needs of the Earth's hydromagnetic dynamo.

On the other hand, Feoktistov nuclear burning progressing wave also sustains another (convective) mechanism of the Earth hydromagnetic dynamo operating. It is explained by the fact that the arrangement of the conditions for gravity convection in the Earth's liquid core, which can be caused by the effective floating up of light fission fragments behind of nuclear burning wave front. The maintenance of continuous fine convection (when temperature is close to adiabatic) in the liquid core is the cause and condition of differential rotation of the different layers of the core , and, hence, the geomagnetic field sustenance.

The mechanism of Feoktistov nuclear-burning progressing wave in combination with the threshold nature of convection in the Earth liquid core can be the reason of the Earth's geomagnetic field inversion. Note that we haven't considered thorium, which in virtue of Eq. (7) can behave as uranium, but with different time scale, i.e. $L/\tau \sim 0.1$ cm/day (where $L \sim 5$ cm is the diffusion distance of neutron in thorium; $\tau$ =39.5/ln2 ≈ 57 days is $^{233}$U generation time due to $\beta$–decay of $^{233}$Pa. This means that the speed of neutron-fission wave propagation in $^{232}$Th-medium (corresponding to Th-U fuel cycle of Teller-Ishikawa-Wood) is one order less than the speed of Feoktistov neutron-fission wave.

Due to the stability of convective currents in Earth's liquid-metal external core has threshold nature, both time scales characterizing the Feoktistov and/or Teller-Ishikawa-Wood progressing wave should set two time scales of the change of convective current stability and, hence, two time scales of the change of the intensity and/or geomagnetic field polarity. It is easy to show that Feoktistov progressing wave goes round the internal core of the Earth in time $\tau_F \sim 1$ million years, whereas this time for Teller-Ishikawa-Wood wave is $\tau_{TIV} \sim 10$ millions years. It is possible to suppose [8] that the polarity change takes place approximately 2 times per hypothetical traversal of the Earth's internal core. Then the crude estimation of the number of geomagnetic polarity changes is

$$2[\tau/\tau_F + \tau/\tau_{TIV} - 2] \sim 170 \qquad (38)$$

and meets well-known Heirtzler scale, in which 171 inversions were fixed during last 79 millions years.

Finally, the marvelous constancy of anomalous isotopic composition of the mantle helium is explained in our case by the properties of fast (~1 MeV) neutron-induced fission of $^{239}$Pu in neutron-fission wave front. The probability of $^3$He production is mainly determined by the probability of $^3$H production as fission fragment of $^{239}$Pu triple fission. This probability is about ~3·10$^{-4}$ [35]. Hence, the total accumulation rate of $^3$He produced as a result of tritium $\beta$–decay (T$_{1/2}$ ~12.3 years) is (with allowance for Eq. (34)) approximately N($^3$He)~10$^{-4}$·$n_f \approx 22.2 \cdot 10^{18}$ $s^{-1}$. On the other hand, the rate of $^4$He accumulation due to $^{238}$U radioactive decay in actinoid web of UO$_2$/Fe equals (with allowance for Eq. (5)) to N($^4$He)~$8H_\alpha/E_\alpha$=8·0.1·10$^{12}$/51.7·1.6·10$^{-13} \approx 9.7 \cdot 10^{22}$ $s^{-1}$. So, in the actinoid web of UO$_2$/Fe (located on the boundary of the solid and liquid phase of the Earth's core) helium ratio is N($^3$He)/N($^4$He) $\approx 2.3 \cdot 10^{-4}$. Obviously that the obtained value is the intermediate value between helium ratio in the Earth's mantle (3·10$^{-5}$) and solar wind (3·10$^{-4}$) and reflects the well-known asymptotics of the relative increase of $^3$He concentration in the depths of the Earth [34].

Thus the hypothesis of slow nuclear burning on the boundary of the liquid and solid phases of the Earth's core is very effective for the explanation of some features of geophysical events. However strong evidences can be obtained from the independent experiment on the measurement of geoantineutrino energy spectrum using the multi-detector scheme of geoantineutrino detection on large base. At the same time the solutions of the direct and inverse problem of neutrino remote diagnostics of the intra-terrestrial processes connected with the obtaining of pure geoantineutrino spectrum and the correct determination of $\beta$–sources radial profile in the Earth's interior will undoubtedly help to solve the problems both of the existence of natural nuclear reactor on the boundary of the liquid and solid phases of the Earth core and true geoantineutrino spectrum.

## Figure captions

Fig. 1. The temporal changes of heat production rate of the Earth within the framework of O'Nionse-Evenson-Hamilton model of geochemical evolution of the mantle differentiation and earth's crust growth (curve U+Th+K), which are supplemented by a nuclear energy source on boundary of solid and liqud phases of the Earth's core (curve U+Th+K+Pu). Crosshatched region is domain of uncertainty.

Fig 2. Concentration kinetic dependences of neutrons (a), $^{238}$U (b); $^{239}$U (c); $^{239}$Pu for different time intervals (d-f), which was obtained by simulation taking into account delayed neutrons ($t$-line is time axis, step $\Delta t=0,01$ s; $x$-line is spatial coordinate axis, step $\Delta x=1$ cm; z-line is concentration axis, particle/cm$^3$). $\Phi_0=1.9\cdot10^{24}$ cm$^{-2}$s$^{-1}$.

Fig. 3. Concentration kinetic dependences of neutrons (a), $^{238}$U (b); $^{238}$U at other visual angle (c), $^{239}$U (d); $^{239}$Pu (e). $\Phi_0=5\cdot10^{17}$ cm$^{-2}$s$^{-1}$.

Fig. 4. The distribution of hot spots on the Earth's surface [8].

Fig. 5. Experimental (•), theoretical double-stochastic Poisson process controlled by response function (o) and Poisson (---)distributions of hot spots on the Earth's surface.

Fig. 6 .Calculated partial antineutrino spectra of $^{239}$Pu normalized on nuclear decay (a) and its deviation from theoratical spectra [24-28] in the energy range $E_{\tilde{\nu}}=1.8$ -10.0 MeV (b).

Fig. 7. Calculated total antineutrino spectrum of the *Earth in KamLAND detector. Firm line is ideal spectrum, histogram is spectrum with the energy resolution of 0.425 MeV.

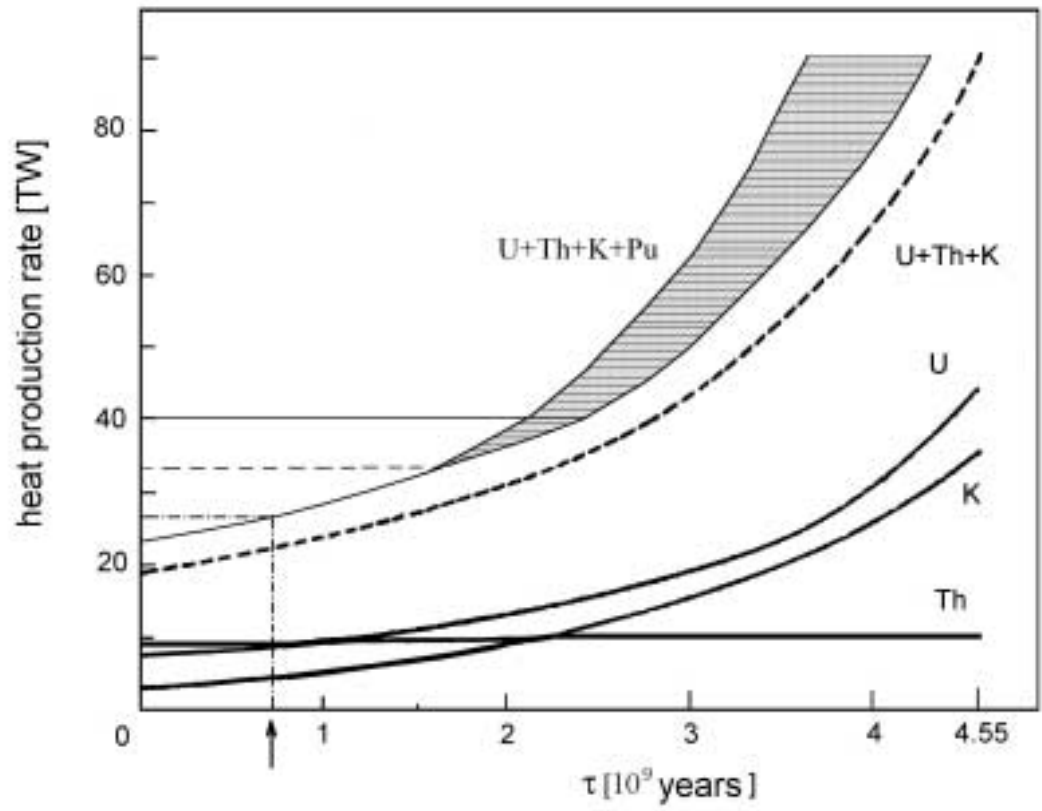

Fig. 1

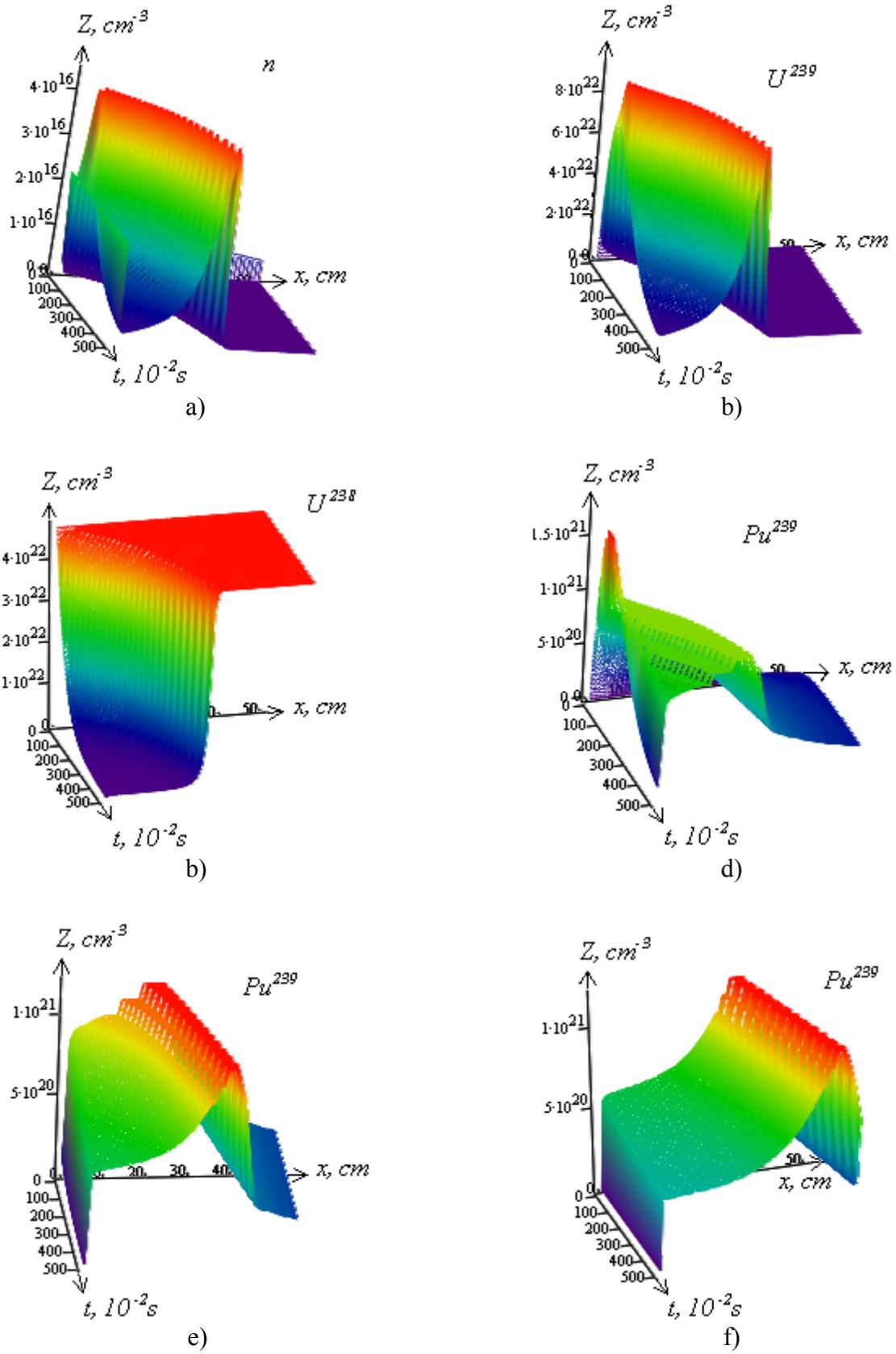

Fig. 2

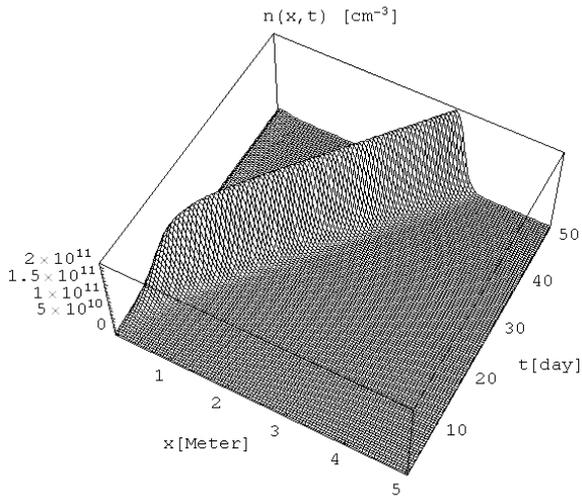

(a)

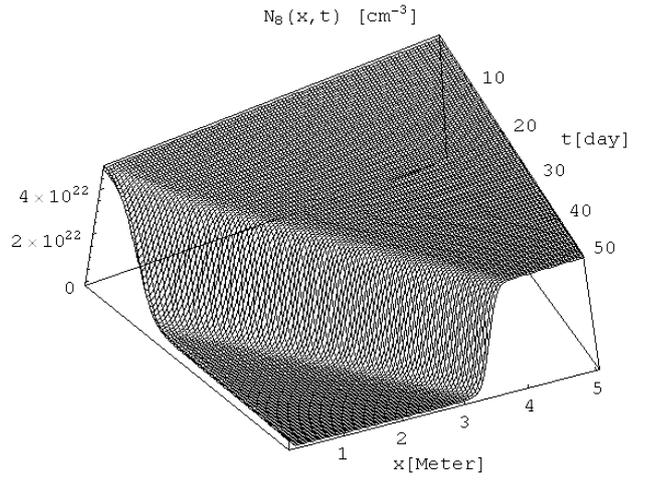

(b)

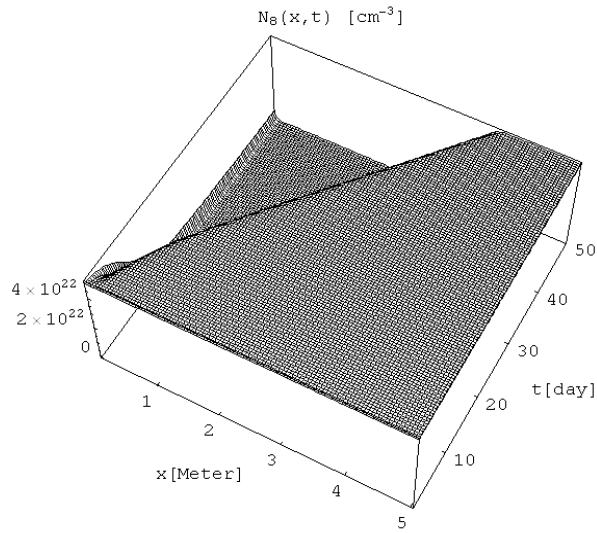

(c)

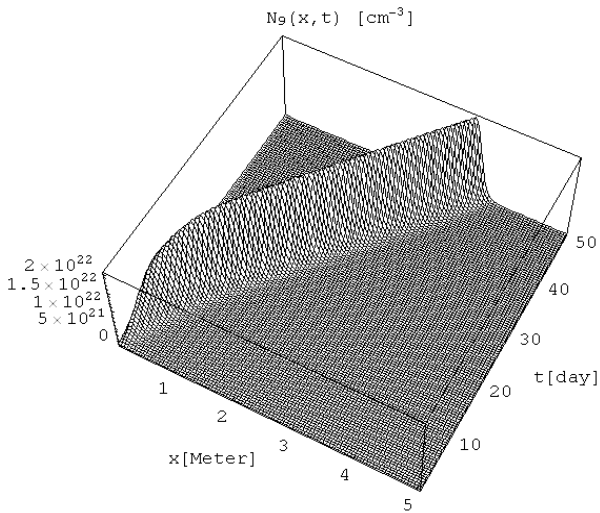

(d)

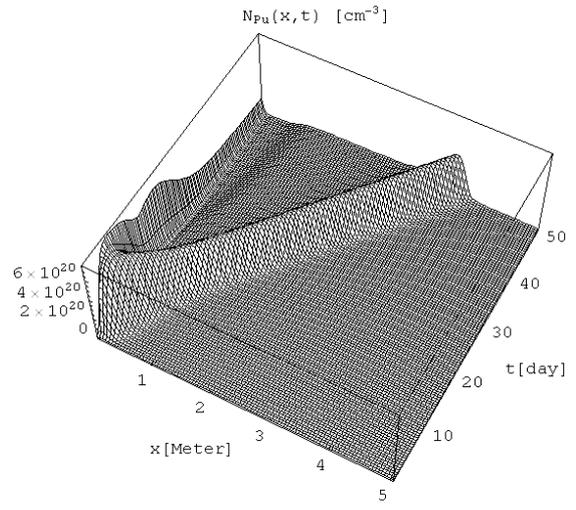

(e)

Fig. 3

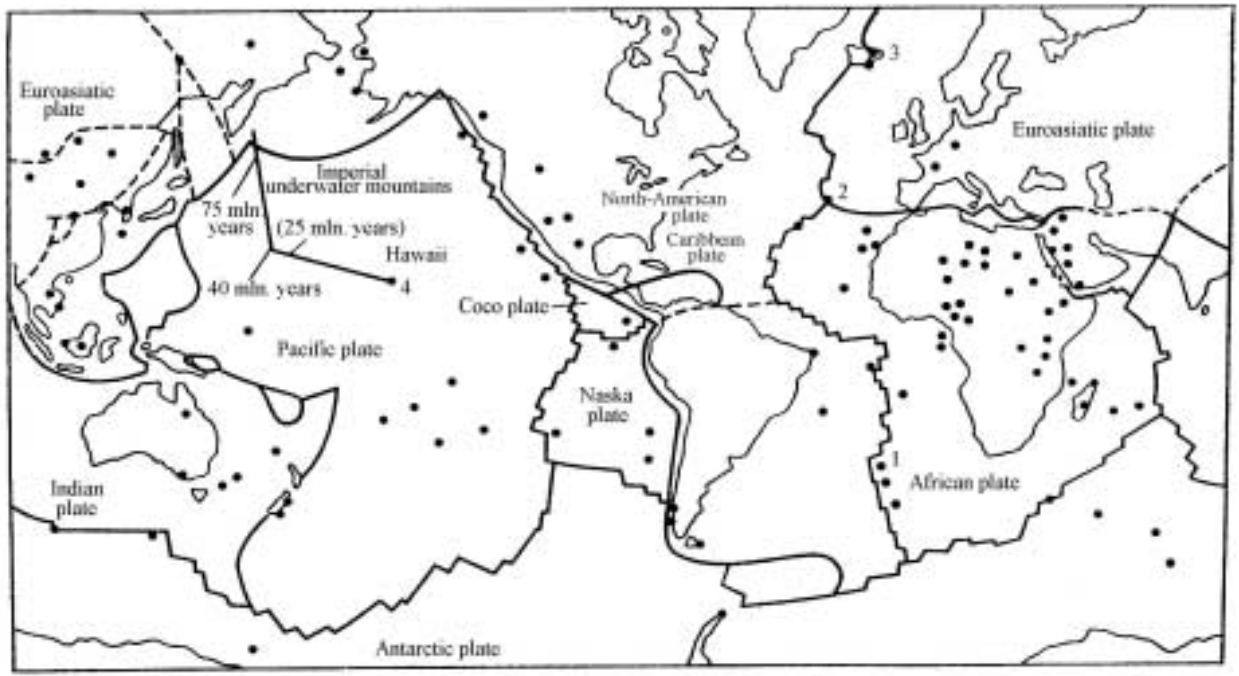

Fig. 4.

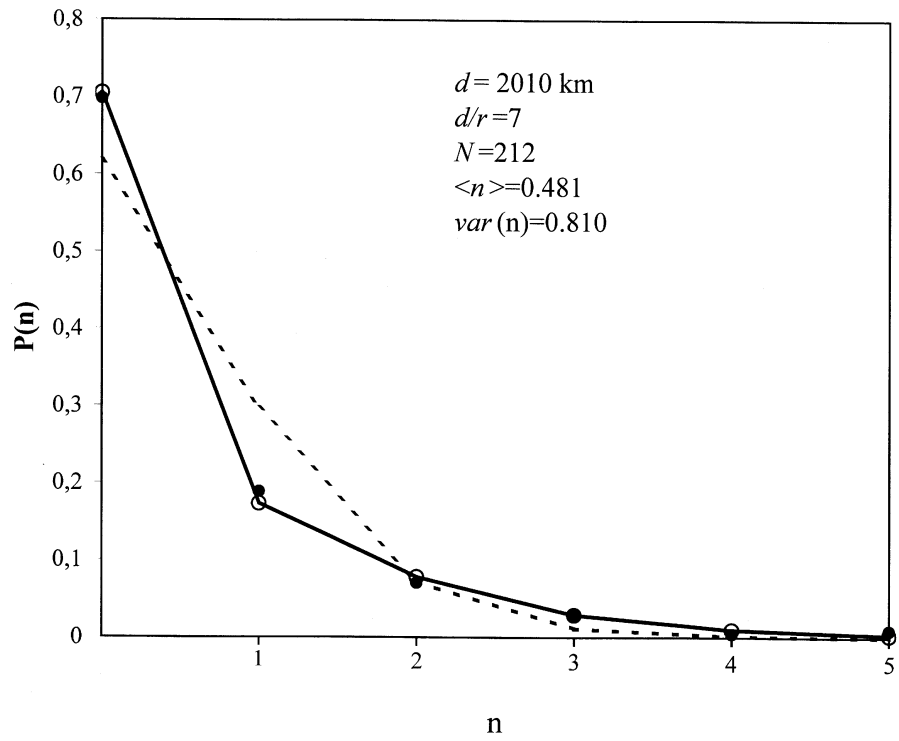

Fig. 5

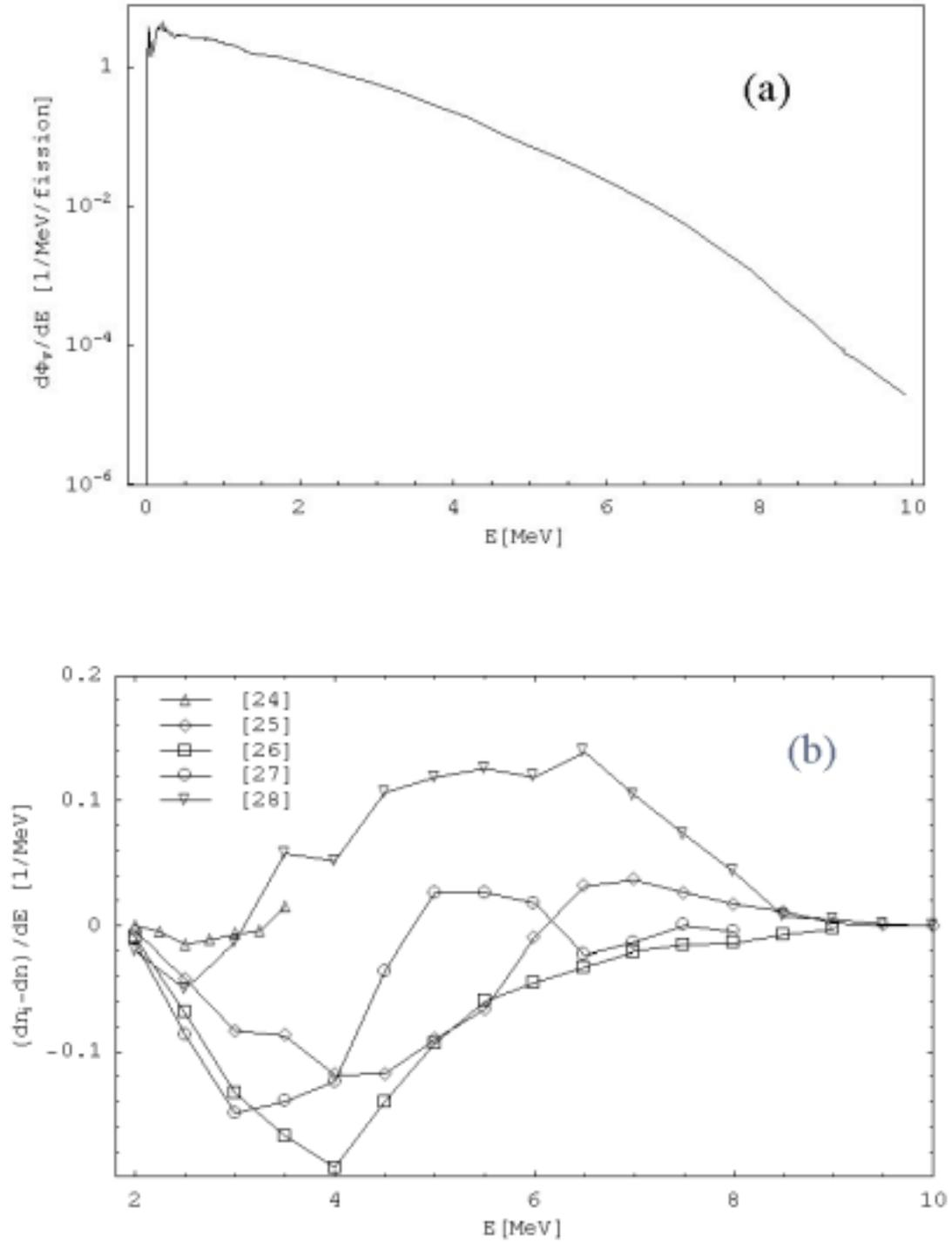

Fig. 6

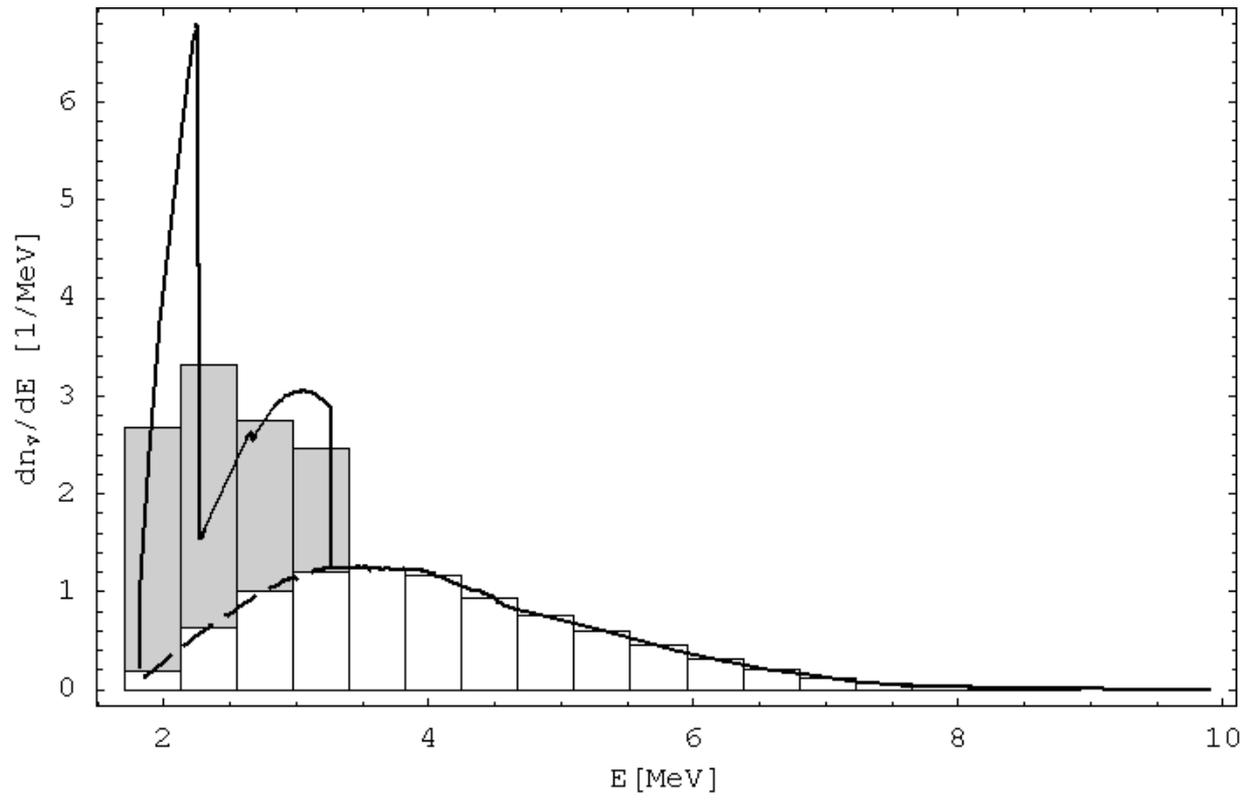

Fig. 7